\documentclass[paper]{JHEP3} 

\usepackage{epsfig,multicol,multirow}
\usepackage{amsmath,subfigure,latexsym,amssymb}

\newcommand\fverb{\setbox\fverbbox=\hbox\bgroup\verb}
\newcommand\fverbdo{\egroup\medskip\noindent%
			\fbox{\unhbox\fverbbox}\ }
\newcommand\fverbit{\egroup\item[\fbox{\unhbox\fverbbox}]}
\newbox\fverbbox

\allowdisplaybreaks

\title{Testing a novel large-$N$ reduction for
${\cal N}=4$ super Yang-Mills theory on $R \times S^3$}

\author{Goro Ishiki,$^{a,b}$ Sang-Woo Kim,$^{c}$
Jun Nishimura$^{b,d}$ and Asato Tsuchiya$^{e}$
\vspace*{0.5cm} \\
\llap{$^a$}Department of Physics, Osaka University,\\
Toyonaka, Osaka 560-0043, Japan\\
\llap{$^b$}High Energy Accelerator Research Organization (KEK),\\
Tsukuba, Ibaraki 305-0801, Japan\\
\llap{$^c$}Center for Quantum Spacetime (CQUeST),\\
Sogang University, Seoul 121-742, Korea\\
\llap{$^d$}Department of Particle and Nuclear Physics,\\
Graduate University for Advanced Studies (SOKENDAI),\\
Tsukuba, Ibaraki 305-0801, Japan\\
\llap{$^e$}Department of Physics, Shizuoka University,\\
836 Ohya, Suruga-ku, Shizuoka 422-8529, Japan
\vspace*{0.5cm} \\
\email{ishiki@post.kek.jp, sangwookim@sogang.ac.kr\\
jnishi@post.kek.jp, satsuch@ipc.shizuoka.ac.jp}}

\preprint{OU-HET 623\\ KEK-TH-1317}

\abstract{Recently a novel large-$N$ reduction has been
proposed as a maximally supersymmetric regularization of
${\cal N}=4$ super Yang-Mills theory
on $R \times S^3$ in the planar limit.
This proposal, if it works, will enable us to study
the theory non-perturbatively on a computer, and hence to
test the AdS/CFT correspondence
analogously to the recent works on the D0-brane system.
We provide a nontrivial check of this proposal
by performing explicit calculations in the large-$N$ reduced model,
which is nothing but the so-called plane wave matrix model,
around a particular stable vacuum
corresponding to $R \times S^3$.
At finite temperature and at weak coupling,
we reproduce precisely the deconfinement phase transition
in the ${\cal N}=4$ super Yang-Mills theory
on $R \times S^3$. This phase transition is considered to
continue to the strongly coupled regime, where it corresponds
to the Hawking-Page transition on the AdS side.
We also perform calculations around other stable vacua,
and reproduce the phase transition in super Yang-Mills theory
on the corresponding curved space-times
such as $R \times S^3/Z_q$ and $R \times S^2$.
}


\begin{document}

\section{Introduction}

The gauge-gravity duality \cite{AdS-CFT} has been one of the
most important subjects in string theory
over the past decade.
The
most typical example is
the so-called AdS/CFT correspondence
between type IIB superstring theory on $AdS_5\times S^5$
and ${\cal N}=4$ U($N$) super Yang-Mills theory (SYM).
Even in this case, however,
a complete proof of the duality is still
missing\footnote{See
refs.\ \cite{kawai-suyama,Berkovits:2007rj}
for some attempts to prove the AdS/CFT correspondence
by using the worldsheet approach.
Also there are remarkable developments based on
the integrability \cite{spin-chain}.}
partly because it is a strong/weak duality.
The region on the string theory side,
where (semi-)classical treatments of gravity or string theory is valid,
is mapped to the strongly
coupled region in the planar large-$N$ limit on the gauge theory side.
In order to study
${\cal N}=4$ SYM
in the strongly coupled regime
from first principles, one needs to have a non-perturbative formulation
such as the lattice gauge theory.
The problem here is that the supersymmetry algebra includes
translational symmetry, which is necessarily broken
by the lattice regularization.
In order to restore supersymmetry in the continuum limit,
one generally has to fine-tune parameters in the lattice action.
In fact there are considerable developments
in reducing the number of parameters to be fine-tuned\footnote{See refs.\
\cite{latticeSUSY,Hanada:2007ti,Catterall:2007fp,AHNT,Catterall:2008yz,%
Hanada:2008gy,Hanada:2008ez}
for recent works.},
but any lattice formulations of ${\cal N}=4$ SYM proposed so far
seem to require fine-tuning at least three parameters \cite{latticeSUSY_N4}.
There are also Monte Carlo studies of the
${\cal N}=4$ SYM based on reduction to matrix quantum mechanics of
6 bosonic commuting matrices
\cite{Berenstein:2008jn,Berenstein:2007wz}, which
confirmed the AdS/CFT correspondence for 1/2 BPS operators.

Here we are aiming at first-principle calculations
in ${\cal N}=4$ U($N$) SYM respecting supersymmetry maximally.
Since we are interested in the planar large-$N$ limit,
we
may well
have a chance to use the idea of
the large-$N$ reduction \cite{EK}.
It asserts that the planar large-$N$ limit of gauge theories
can be studied by dimensionally reduced models, which can be
obtained by dimensional reduction.
The original idea
does not work
in general because of the
spontaneous breaking of the U(1)$^D$ symmetry in the reduced
model \cite{Bhanot},
which led to various
proposals \cite{Bhanot,Parisi,Gross,Das,GonzalezArroyo,%
Narayanan:2003fc,Kovtun:2007py}.
Since reduced models can be regularized by making $N$ finite,
one may avoid introducing the lattice structure in space-time \cite{Gross},
which causes the breaking of supersymmetry.
However, it seemed rather difficult to avoid the problem concerning
the instability of the U(1)$^D$ symmetric
vacuum
without breaking supersymmetry.

Let us recall here that ${\cal N}=4$ U($N$) SYM becomes conformally
invariant
if one sets all the moduli parameters
(represented by the expectation values of adjoint scalars)
to zero.
Hence the theory on $R^4$ at the conformally invariant
point in the moduli space
is equivalent to the theory on $R\times S^3$ through
conformal mapping.
In fact,
$R\times S^3$ is obtained as the boundary of
$AdS_5$, when one uses the global coordinate.
Thus $\mathcal{N}=4$ SYM on $R\times S^3$
appears naturally in the context of
the AdS/CFT correspondence,
for instance,
in the so-called pp-wave
limit \cite{Berenstein:2002jq,Berenstein:2002sa,Okuyama:2002zn}
and in the bubbling AdS \cite{Lin:2004nb,Corley:2001zk,Berenstein:2004kk}.

For our purpose it is intriguing to dimensionally
reduce the theory by collapsing the $S^3$ of
$R\times S^3$ to a point.
The one-dimensional gauge theory obtained in this way
is nothing but
the plane wave matrix model (PWMM) \cite{Berenstein:2002jq}
as pointed out by ref.\ \cite{Kim-Klose-Plefka}.\footnote{In
ref.\ \cite{Kim-Klose-Plefka}
an equivalence between PWMM around the
\emph{trivial vacuum}
and the \emph{pp-wave limit}
of $\mathcal{N}=4$ SYM on $R\times S^3$ has been shown
at one loop in the pure scalar sector.
However, discrepancies are found by four-loop
calculations \cite{Fischbacher:2004iu}.
This connection between PWMM and SYM should not be confused
with the large-$N$ reduction we are going to discuss.}
The PWMM can be regarded as a mass-deformation
of the Matrix theory \cite{BFSS} preserving maximal supersymmetry,
where the mass parameter corresponds to the curvature of the
$S^3$ before dimensional reduction.\footnote{Originally
PWMM appeared as a generalization of the Matrix theory
to the pp-wave background \cite{Berenstein:2002jq}.
It is also often referred to as the BMN matrix model in the literature.
While we obtain formally the same model
in the context of large-$N$ reduction, the interpretation
of the model is different.
}
The model possesses many classical vacua,
all of which preserve maximal supersymmetry.

Recently it has been conjectured \cite{Ishii:2008ib}
that if one picks up a particular
classical vacuum of the PWMM,
which corresponds to a sequence of
fuzzy spheres
with different radii,
one can actually retrieve the theory
before dimensional reduction in the planar limit.
(See refs.\ \cite{Ishiki:2006yr,Ishii:2007ex,Ishii:2008tm}
for earlier discussions.)
%
%
%
%
This conjecture may be
viewed as
a new type of large-$N$ reduction, which
extends the original proposal for the flat space-time
to a curved one, and at the same time
solves the aforementioned problems
concerning the
vacuum instability.
The classical instability is avoided
since the PWMM is a massive theory,
and the quantum instability is avoided, too, since
the
vacuum preserves
maximal supersymmetry.
Since the planar limit is taken in the reduced model,
the instanton transition to other vacua
is also suppressed and the ``fuzziness'' of the spheres
is removed.

Viewed as a regularization
of the ${\cal N}=4$ SYM on $R\times S^3$,
the present formulation respects the maximal
${\rm SU}(2|4)$ supersymmetry (with 16 supercharges)
of the PWMM, and in the large-$N$ limit
the symmetry is expected to enhance to the full superconformal
${\rm SU}(2,2|4)$ symmetry,
which has 32 supercharges. Considering that
the conformal symmetry is
broken by any kind of UV regularizations, this regularization
is optimal from the viewpoint of preserving supersymmetries.

Let us also emphasize that
if one naively regularizes the
${\cal N}=4$ SYM on $R\times S^3$
by introducing an upper bound
on the angular momenta on $S^3$,
one necessarily breaks the gauge symmetry as well as
supersymmetry. This problem can be dealt with in perturbative
calculations by adding appropriate
counter-terms \cite{Aharony:2005bq,Ishiki:2006rt},
but it is not clear what to do with it
in nonperturbative calculations that we are aiming at eventually.
In
the present formulation,
the size of the matrices plays the role of
the ultraviolet cutoff, which
neatly
respects both gauge symmetry and supersymmetry.

In this paper
we test the novel large-$N$ reduction at weak coupling.
In supersymmetric theories, it often occurs that
certain properties in the strongly coupled regime
remain qualitatively the same
in the weakly coupled regime.
For instance,
the AdS/CFT correspondence at finite temperature \cite{Witten:1998zw}
suggests that
there is a first-order phase transition in the strongly coupled regime
of ${\cal N}=4$ SYM on $R\times S^3$ in the planar limit,
which corresponds, on the gravity side,
to the Hawking-Page transition \cite{Hawking:1982dh}
between the AdS space-time and the AdS black hole.
In fact, even in the weak coupling limit of the ${\cal N}=4$ SYM,
there exists a first-order deconfinement
phase
transition \cite{Sundborg:1999ue,Aharony:2003sx},
which is conjectured to be a continuation of the one in the
strongly coupled regime.\footnote{Thermodynamical properties of
${\cal N}=4$ SYM on $R\times S^3$ have been
studied also in refs.\ \cite{Yamada:2006rx,Harmark:2007px}}
We confirm the novel large-$N$ reduction
at weak coupling by showing
that the PWMM indeed reproduces precisely
the above phase transition.
The main results of this work were reported briefly
in our previous publication \cite{Ishiki:2008te}.
As a related work,
a test of the large-$N$ reduction has been performed
in the high temperature limit up to two-loop \cite{Kitazawa:2008mx}
(See also ref.\ \cite{Kaneko:2007ui}.).
An application of the large-$N$ reduction
to ${\cal N}=1$ SYM on $R\times S^3$ was
discussed in ref.\ \cite{Hanada:2009kz}.

This paper is organized as follows.
In section 2 we briefly review the deconfinement phase transition
in ${\cal N}=4$ SYM on $R\times S^3$ in the weak coupling limit.
In section 3 we describe the large-$N$ reduction
proposed in ref.\ \cite{Ishii:2008ib}.
In section 4 we discuss
the weak coupling limit of
the PWMM around the
vacuum
corresponding to $R \times S^3$
at finite temperature.
In section 5 we show analytically that the critical temperature of
the PWMM
agrees with that of ${\cal N}=4$ SYM on $R\times S^3$.
Derivation of some equations is given in appendix \ref{sec:appendixA}.
In section 6 we show analytically
that the free energy of
the PWMM
agrees with that of ${\cal N}=4$ SYM on $R\times S^3$
under some assumption.
In section 7 we perform Monte Carlo simulations
to verify this assumption and
to demonstrate the agreement of the free energy explicitly.
In section 8 we show that ${\cal N}=4$ SYM on
a more general space-time $R \times S^3/Z_q$ can be obtained
by choosing a different classical vacuum of the same model.
In appendix \ref{sec:appendixB}
we discuss a simpler case of
${\cal N}=8$ SYM on $R \times S^2$, which
can be obtained from coinciding fuzzy spheres.
Section 9 is devoted to a summary and discussions.

\section{Brief review of weakly coupled ${\cal N}=4$ SYM on $R\times S^3$}
\label{section:review}

In this section we briefly review
the calculation \cite{Sundborg:1999ue,Aharony:2003sx},
which showed that weakly coupled
${\cal N}=4$ U($k$) SYM on $R \times S^3$
undergoes a deconfinement phase transition
in the planar large-$k$ limit at finite temperature.
We will see later that equations similar to the ones
that appear below are reproduced from
the reduced model.
In order to make the similarity clearer,
we use $k$ instead of $N$
for
the gauge group in this section.

Let us introduce a finite temperature $T$ by
compactifying the Euclidean time $t$
to a circle with the circumference $T^{-1}$.
Unlike the $T=0$ case, the holonomy along the $t$ direction
becomes nontrivial and it is
represented by
the holonomy matrix $U$.
One can choose a gauge, in which $U$ takes the diagonal form
\begin{align}
U=\mbox{diag}(e^{i\alpha_1},\cdots,e^{i\alpha_k}) \ ,
\end{align}
where $\alpha_{a} \in (-\pi , \pi]$ ($a=1, \cdots , k$)
are constant in space-time.
The $\alpha_a$ variables
are called the gauge field moduli,
since they are
massless zero modes.

In the weak coupling limit,
all the fields except the gauge field moduli
can be integrated out at one loop.
Since we are going to take the large-$k$ limit,
it is convenient to introduce the distribution
of the gauge field moduli
\begin{align}
\rho(\theta)=\frac{1}{k}\sum_{a=1}^k\delta(\theta-\alpha_a)\ .
\end{align}
The resulting effective theory for $\rho(\theta)$
is given by \cite{Sundborg:1999ue,Aharony:2003sx}
\begin{align}
&S=k^2\int d\theta d\theta' \,
\rho(\theta) \, V(\theta-\theta') \, \rho(\theta') \ ,
\label{continuum effective action}\\
\label{Vthetadef}
&V(\theta) =
\sum_{p=1}^{\infty}\tilde{V}_p \, \cos(p\theta) \  , \\
&\tilde{V}_p =
\frac{1}{p}\Bigl\{1-6z_s(x^p)-z_v(x^p)
-4(-1)^{p+1}z_f(x^p) \Bigr\} \ ,
\label{Vdef}
\end{align}
where we have introduced dimensionless parameters
\begin{align}
\label{defx}
 x = e^{ - \beta } \ , \quad
 \beta =
\frac{1}{R_{S^3} T}
\end{align}
with $R_{S^3}$ being the radius of $S^3$.
We have also introduced the functions
\begin{align}
z_s(x)=\frac{x+x^2}{(1-x)^3} \ , \quad
z_v(x)=\frac{6x^2-2x^3}{(1-x)^3} \ , \quad
z_f(x)= \frac{4x^{\frac{3}{2}}}{(1-x)^3} \ ,
\label{single-particle partition function}
\end{align}
which can be interpreted as
the single-particle partition functions
for the scalars, the vector and the fermions, respectively.
Then one can obtain the distribution
$\rho(\theta)$ exactly in the large-$k$ limit
by solving the saddle-point equation
\begin{align}
\int_{-\pi}^{\pi}d\theta' \, V'(\theta-\theta') \, \rho(\theta')
= 0 \ .
\label{saddle-point equation for continuum theory}
\end{align}

Obviously the uniform distribution is always a solution to
the saddle-point equation.
At low temperature, it gives the absolute minimum of
the effective action.
As a consequence, the center invariance is unbroken and
the free energy normalized by $\frac{1}{k^2}$,
which we call the \emph{normalized free energy} in what follows,
vanishes in the large-$k$ limit.
This phase can be interpreted as the confined phase.
One can show that there is a first order phase transition
at a critical point determined by
\begin{align}
\tilde{V}_1=0 \ ,
\label{condition}
\end{align}
which gives $x_c=7-4\sqrt{3}$ \cite{Sundborg:1999ue,Aharony:2003sx}
in terms of the dimensionless parameter (\ref{defx}).
%
%
%

Above the critical temperature,
the dominant solution has a compact support
$[-\theta_0,\theta_0]$ with $\theta_0<\pi$,
and the equation
(\ref{saddle-point equation for continuum theory})
is satisfied only for $\theta \in [-\theta_0,\theta_0]$.
The center invariance is broken and
the normalized free energy takes a negative value.
This phase can be interpreted as the deconfined phase.
Near the critical temperature, in particular,
the explicit form of $\rho(\theta)$ is
given by the Gross-Witten form
\cite{Sundborg:1999ue,Aharony:2003sx}
\begin{align}
& \rho (\theta)=\left\{
\begin{array}{ll}
\frac{1}{\pi \omega}
\left(\cos \frac{\theta}{2}\right)
\sqrt{ \omega-\sin^2 \frac{\theta}{2} }   &
\;\mbox{for} \;\;|\theta|\leq \theta_0 \\
0 & \;\mbox{for}\;\; |\theta| > \theta_0 \ ,
\end{array}  \right.
\label{eigen_anal} \\
& \mbox{where} \;\;
\theta_0=2 \sin ^{-1} \sqrt{\omega}\ , \quad
\omega =
1-\sqrt{\frac{-\tilde{V}_1}{1-\tilde{V}_1}} \ .
\end{align}
The normalized free energy above the critical temperature
is obtained for
$R_{S^3}=1$
as \cite{Aharony:2003sx}
\begin{align}
\frac{F_{\rm SYM}}{k^2}
=-0.9877(T-T_c)-4.248(T-T_c)^{\frac{3}{2}}
-11.696(T-T_c)^2+{\cal O}((T-T_c)^{\frac{5}{2}}) \ ,
\label{F_SYM}
\end{align}
where $T_c=-1/\ln(7-4\sqrt{3})=0.37966\cdots$.
This phase transition is
speculated to be
a continuation of the conjectured phase transition at strong coupling,
which corresponds to the Hawking-Page transition \cite{Hawking:1982dh}
according to the AdS/CFT correspondence \cite{Witten:1998zw}.

\section{Large-$N$ reduction for
${\cal N}=4$ SYM on $R\times S^3$}

In order to regularize
${\cal N}=4$ U($N$) SYM on $R\times S^3$
respecting supersymmetry maximally,
we use the idea of the large-$N$ reduction.
%
For that we dimensionally reduce the theory
by collapsing the $S^3$
to a point.
Thus we obtain
a one-dimensional gauge theory
\begin{align}
S_{\rm PWMM}
= \frac{1}{g
^2}
\int
dt \, \mbox{tr} &
\left[\frac{1}{2}(D_tX_M)^2-\frac{1}{4}[X_M,X_N]^2
+\frac{1}{2}\Psi^{\dagger} D_t \Psi
-\frac{1}{2}\Psi^{\dagger}\gamma_M[X_M,\Psi] \right.\nonumber\\
&
\left.+\frac{\mu^2}{2}(X_i)^2
+\frac{\mu^2}{8}(X_a)^2 +i\mu\epsilon_{ijk}X_iX_jX_k
+i\frac{3\mu}{8}\Psi^{\dagger}\gamma_{123}\Psi \right] \ ,
\label{pp-action}
\end{align}
which is
nothing but the PWMM \cite{Berenstein:2002jq}.
Here the covariant derivative is defined by
$D_t=\partial_t-i[A, \ \cdot \ ]$,
where $A$, as well as $X_M$ and $\Psi$, is
an $N\times N$ matrix depending on $t$.
The range of indices is given by
$1 \le M,N \le 9$, $1 \le i,j,k \le 3$ and $4 \le a \le 9$.
The model has SU$(2|4)$ supersymmetry, which includes
16 supercharges,
and for $\mu=0$ it reduces to the D0-brane effective theory
or the Matrix theory \cite{BFSS}.

In fact the model possesses many vacua
representing multi fuzzy spheres.
Explicitly, they are given by
\begin{align}
X_i=\mu \bigoplus_{I=1}^{\nu}
\Bigl( L_i^{(n_I)}\otimes {\bf 1}_{k_I} \Bigr) \ ,
 \label{background}
\end{align}
where $L_i^{(n)}$ are
the $n$-dimensional irreducible representation of
the SU$(2)$ generators obeying
$[L_i^{(n)},L_j^{(n)}]=i \, \epsilon_{ijk} \, L_k^{(n)}$.
The parameters $n_I$ and $k_I$ in (\ref{background})
have to satisfy the relation $\sum_{I=1}^{\nu}n_Ik_I=N$.
All of these vacua preserve the SU$(2|4)$ supersymmetry,
and they are degenerate.

In order to retrieve ${\cal N}=4$ SYM on $R \times S^3$
in the planar limit, one
has to pick up a particular background
from (\ref{background}),
and consider the theory (\ref{pp-action}) around it.
Let us consider the case
\begin{align}
k_I=k \ , \quad n_I=n+I-\frac{\nu+1}{2} \quad \quad
\mbox{for $I =1, \cdots , \nu$}
\label{our background}
\end{align}
with odd $\nu$,
and take the large-$N$ limit in such a way that
\begin{align}
&n \rightarrow\infty, \;\;
\nu\rightarrow\infty, \;\;k\rightarrow \infty,\;\;
n-\frac{\nu}{2}\rightarrow\infty\nonumber\\
&\mbox{with} \;\;
\lambda \equiv \frac{g^2 k}{n}
\; \; \mbox{fixed} \ .
\label{limit}
\end{align}
%
%
Then the resulting theory is claimed
\cite{Ishii:2008ib} to be equivalent
to the planar limit of
${\cal N}=4$ SYM on $R \times S^3$
with the radius of $S^3$
and the 't Hooft coupling constant given,
respectively, by\footnote{The relationship between
the radius of $S^3$
and the parameter $\mu$ of the PWMM
agrees with the one obtained in
dimensionally reducing
${\cal N}=4$ SYM on $R\times S^3$ to arrive at
the PWMM (\ref{pp-action}).\label{foot:radiusS3}}
\begin{align}
R_{S^3}=\frac{2}{\mu} \ ,
\label{defRS3}  \quad \quad
 \lambda_{R \times S^3} = \lambda \, V_{S^3} \ ,
\end{align}
where $V_{S^3}= 2\pi^2 (R_{S^3})^3$ is the volume of $S^3$.

This equivalence may be viewed as an extension
of the Eguchi-Kawai equivalence \cite{EK} to a curved space-time.
It is crucial that we do not need to do anything like
momentum quenching \cite{Bhanot,Gross}.
As a consequence, the formulation
preserves the ${\rm SU}(2|4)$ supersymmetry and the gauge symmetry.
For a brief review of the equivalence, see ref.\ \cite{Ishiki:2008te}.
In the following sections, we give a nontrivial test of
this proposal by studying the model at weak coupling.

\section{Effective theory for the gauge field moduli}

Let us first consider a perturbative expansion
of the PWMM
around
the most general background (\ref{background}).
The calculation is analogous to what is done
in ${\cal N}=4$ SYM on $R\times S^3$ in section \ref{section:review}.
In particular, we introduce finite temperature
$T$ in (\ref{pp-action}),
and choose a gauge in which
the holonomy matrix $U$
is diagonal
\begin{align}
 U =& \bigoplus_{I=1}^{\nu}
\Bigl( {\bf 1}_{n_I} \otimes U_I \Bigr) \ , \\
 U_{I}=& \mbox{diag}
\Bigl(e^{i\alpha^{(I)}_1},
\cdots,e^{i\alpha^{(I)}_{k_I}}\Bigr)
\quad\quad \mbox{for $I=1,\cdots,\nu$} \ ,
\end{align}
where $\alpha_{a}^{(I)} \in (-\pi, \pi] \;(a=1,\cdots,k_I)$.
In what follows we use the dimensionless
parameters $\beta$ and $x$ as defined
in eq.\ (\ref{defx})
with $R_{S^3} = 2/\mu$ anticipating the relationship
(\ref{defRS3}).

The effective action for the gauge field moduli
around the general background (\ref{background}) is obtained
in the weak coupling limit \cite{Kawahara:2006hs}.
It can be decomposed into
\begin{align}
S_{\rm eff}=6S_s+S_v+4S_f+S_V  \ ,
\label{effective action for general background}
\end{align}
where $S_s$, $S_v$, $S_f$ and $S_V$ represent
the contribution of a scalar, a vector, a fermion and
the Vandermonde determinant, respectively.
Explicitly, they are given as
\begin{align}
&S_s=\sum_{I,J=1}^{\nu} S_s^{(I,J)} \ , \quad
S_v=\sum_{I,J=1}^{\nu} S_v^{(I,J)}\ , \quad
S_f=\sum_{I,J=1}^{\nu} S_f^{(I,J)}\ , \quad
S_V=\sum_{I=1}^{\nu}S_V^{(I)}\ ,
\label{svfV} \\
\nonumber\\
&S_s^{(I,J)}
=\frac{1}{2}\sum_{a=1}^{k_I}\sum_{b=1}^{k_J}
\sum_{l=|n_I-n_J|/2}^{(n_I+n_J)/2-1} \!\!\!\!\!\!\! (2l+1)
\ln \Bigl\{ 1+e^{-2\beta (2l+1)}-2e^{-\beta (2l+1)}
\cos (\alpha_a^{(I)}-\alpha_b^{(J)}) \Bigr\} \ ,
\label{scalar} \\
&S_v^{(I,J)}
=\frac{1}{2} \sum_{a=1}^{k_I}\sum_{b=1}^{k_J}
\left[\sum_{l=|n_I-n_J|/2-1+\delta_{IJ}}^{(n_I+n_J)/2-2}
\!\!\!\!\!\!\! (2l+1)\ln
\Bigl\{ 1+e^{-2\beta(2l+2)}-2e^{-\beta(2l+2)}
\cos(\alpha_a^{(I)}-\alpha_b^{(J)}) \Bigr\}
\right.\nonumber\\
&\left.\qquad\qquad
+\sum_{l=|n_I-n_J|/2+1}^{(n_I+n_J)/2}\!\!\!\!\!\!\!(2l+1)\ln
\Bigl\{ 1+e^{-2\beta \cdot 2l}-2e^{-\beta \cdot 2l}
\cos(\alpha_a^{(I)}-\alpha_b^{(J)}) \Bigr\} \right]\ ,
\label{vector} \\
&S_f^{(I,J)}
=- \frac{1}{2}\sum_{a=1}^{k_I}\sum_{b=1}^{k_J}
\left[\sum_{l=|n_I-n_J|/2-1/2}^{(n_I+n_J)/2-3/2}\!\!\!\!\!\!\!(2l+1)\ln
\Bigl\{ 1+e^{-2\beta (2l+\frac{3}{2})}+2e^{-\beta (2l+\frac{3}{2})}
\cos(\alpha_a^{(I)}-\alpha_b^{(J)}) \Bigr\}
\right.\nonumber\\
&\left.
\qquad\qquad
+ \sum_{l=|n_I-n_J|/2+1/2}^{(n_I+n_J)/2-1/2}\!\!\!\!\!\!\!(2l+1)\ln
\Bigl\{ 1+e^{-2\beta (2l+\frac{1}{2})}+2e^{-\beta (2l+\frac{1}{2})}
\cos(\alpha_a^{(I)}-\alpha_b^{(J)}) \Bigr\} \right] \ ,
\label{fermion} \\
&S_V^{(I)}=-\sum_{a\neq b}^{k_I}
\ln\left|\sin\frac{\alpha_a^{(I)}-\alpha_b^{(I)}}{2}\right|
 -(k_I^2-k_I)\ln2\ .
\label{Vandermonde}
\end{align}
The second term in (\ref{Vandermonde})
is needed to make the free energy vanish in the low temperature
phase. With this constant term,
the total effective action
(\ref{effective action for general background})
agrees exactly with the one presented in ref.\ \cite{Kawahara:2006hs}.

Let us restrict ourselves here
to the particular case (\ref{our background}), and rewrite
the effective action (\ref{effective action for general background})
in a form analogous to
(\ref{continuum effective action})$\sim$(\ref{Vdef}),
which is useful for analytical studies.
For instance, $S_s^{(I,J)}$ in (\ref{scalar})
can be rewritten as
\begin{align}
S_s^{(I,J)}&=\frac{1}{2}\sum_{a,b=1}^k
\sum_{l=|n_I-n_J|/2}^{(n_I+n_J)/2-1}(2l+1)
\Biggl[ \ln \Bigl\{ 1-e^{-\beta(2l+1)+
i(\alpha_a^{(I)}-\alpha_b^{(J)})} \Bigr\}
+ \mbox{c.c.} \Biggr]
\nonumber \\
&=-\sum_{p=1}^{\infty}\frac{1}{p}
\sum_{a,b=1}^k\cos \Bigl\{ p(\alpha_a^{(I)}-\alpha_b^{(J)}) \Bigr\}
\sum_{l=|n_I-n_J|/2}^{(n_I+n_J)/2-1}(2l+1)e^{-p\beta(2l+1)} \nonumber\\
&=\sum_{p=1}^{\infty}\frac{1}{p^2}
\sum_{a,b=1}^k\cos \Bigl\{ p(\alpha_a^{(I)}-\alpha_b^{(J)}) \Bigr\}
\frac{\partial}{\partial \beta}
\left(\frac{e^{-p\beta(|n_I-n_J|+1)}
(1-e^{-p\beta (n_I+n_J-|n_I-n_J|)})}{1-e^{-2p\beta}}\right) \ .
\nonumber
\end{align}
We also introduce the distribution function for $\alpha^{(I)}_a$ as
\begin{align}
\rho^{(I)}(\theta)=
\frac{1}{k}\sum_{a=1}^k\delta(\theta-\alpha_a^{(I)}) \quad\quad
\mbox{for $I=1,\cdots,\nu$} \ .
\end{align}
Then we find that the effective action for the gauge field moduli can
be written as
\begin{align}
 S_{\rm eff}  =&  k^2 \!\! \sum_{I,J=1}^{\nu}
\int
d\theta d\theta' \,
\rho^{(I)}(\theta)
\, V^{(I,J)}(\theta-\theta') \,
\rho^{(J)}(\theta') \ ,
\label{effective action for matrix model} \\
 V^{(I,J)}(\theta) =&
\sum_{p=1}^{\infty} \tilde{V}^{(I,J)}_p\cos(p\theta) \ ,
\label{Vij-def}
\\
 \tilde{V}^{(I,J)}_p =&
\frac{1}{p}
\Bigl\{
\delta_{IJ}-6z^{(I,J)}_{s}(x^p)-z^{(I,J)}_{v}(x^p)
 -4(-1)^{p+1}z^{(I,J)}_{f}(x^p) \Bigr\} \ .
\label{Vtilde}
\end{align}
We have introduced
the functions
\begin{align}
z_{s}^{(I,J)}(x)
=&x\frac{\partial}{\partial x}\left(
\frac{x^{|n_I-n_J|+1}(1-x^{n_{IJ}})}{1-x^2}\right) \ ,
\label{zsp} \\
z_{v}^{(I,J)}(x)=&x^2\frac{\partial}{\partial x}
\left(\frac{x^{|n_I-n_J|-1+2\delta_{IJ}}(1-x^{n_{IJ}-2\delta_{IJ}})}
{1-x^2}\right)
+\frac{\partial}{\partial x}
\left(\frac{x^{|n_I-n_J|+3}(1-x^{n_{IJ}})}
{1-x^2}\right) \ ,
\label{zvp} \\
z_{f}^{(I,J)}(x)=&x^{\frac{3}{2}}\frac{\partial}{\partial x}
\left(\frac{x^{|n_I-n_J|+2\delta_{IJ}}(1-x^{n_{IJ}-2\delta_{IJ}})}
{1-x^2}\right)
+x^{\frac{1}{2}}\frac{\partial}{\partial x}
\left(\frac{x^{|n_I-n_J|+2}(1-x^{n_{IJ}})}
{1-x^2}\right)
\label{zfp}
\end{align}
with $n_{IJ}= n_I+n_J-|n_I-n_J|$,
which can be interpreted as
the single-particle partition functions
for the scalars, the vector and the fermions, respectively,
as in the $\mathcal{N}=4$ SYM.
We can analyze the effective
theory (\ref{effective action for matrix model})
by the saddle-point method
in the large-$k$ limit.
The free energy is given by
\begin{align}
F_{\rm PWMM} = - T S_{\rm eff} \ ,
\label{def-free-energy}
\end{align}
where $S_{\rm eff}$ is evaluated at the dominant saddle point.
We will show in section \ref{section:free-energy} that
\begin{align}
\frac{1}{k^2 \nu} F_{\rm PWMM} = \frac{1}{k^2} F_{\rm SYM}
\label{agreement of free energy}
\end{align}
in the limit (\ref{limit}), where the right-hand side
is the normalized free energy of the $\mathcal{N}=4$ U($k$) SYM
on $R \times S^3$.
Therefore, we define the normalized free energy for the
PWMM by the left-hand side of eq.\ (\ref{agreement of free energy}).
The appearance of the $\frac{1}{\nu}$ factor
in
the above
relation is consistent with diagrammatic considerations \cite{Ishii:2008ib}.

Since $\rho^{(I)}(\theta)=\rho^{(I)}(-\theta)$
due to symmetry, $\rho^{(I)}(\theta)$ can be expanded as
\begin{align}
\rho^{(I)}(\theta)=
\frac{1}{2\pi}+\frac{1}{\pi}\sum_{p=1}^{\infty}
\tilde{\rho}^{(I)}_p\cos(p\theta) \ .
\end{align}
In terms of $\tilde{\rho}^{(I)}_p$,
the effective action (\ref{effective action for matrix model})
is expressed as
\begin{align}
S_{\rm eff}=
k^2 \!\! \sum_{I,J=1}^{\nu}\sum_{p=1}^{\infty}
\tilde{\rho}^{(I)}_p \, \tilde{V}^{(I,J)}_p \, \tilde{\rho}^{(J)}_p \ .
\label{eff-actionPWMM}
\end{align}
In the low temperature regime,
$\tilde{V}^{(I,J)}_p$
given by (\ref{Vtilde})
are positive definite matrices.
The action (\ref{eff-actionPWMM})
is therefore minimized by
a configuration with $\rho^{(I)}_p=0$ for all $p\geq 1$,
which implies that $\alpha^{(I)}_a$ distribute uniformly.
The center invariance is unbroken and
the normalized free energy
vanishes.
This phase can be interpreted as the confined phase.
Above the temperature determined by
\begin{align}
\det \tilde{V}^{(I,J)}_p=0 \;\;\; \mbox{for certain} \;\; p \ ,
\label{condition for critical temperatute}
\end{align}
we obtain a configuration with $\rho^{(I)}_p\neq 0$
for some $I$ and $p\geq 1$,
which implies that
some of the distributions $\rho^{(I)}(\theta)$ become non-uniform.
As a consequence, the center invariance is broken,
and the normalized free energy becomes negative.
The high temperature phase can be interpreted as the deconfined phase.
The distributions $\rho^{(I)}(\theta)$ are determined
by the saddle-point equation
\begin{align}
\sum_{J=1}^{\nu}\int
d\theta' V^{(I,J)}(\theta-\theta') \rho^{(J)}(\theta')=0
\;\;\; \mbox{for} \;\; \theta \in [-\theta_0^{(I)},\theta_0^{(I)}]
\label{saddle-point equation for matrix model}
\end{align}
derived from (\ref{effective action for matrix model}),
where
$[-\theta_0^{(I)},\theta_0^{(I)}]\;\; (\theta_0^{(I)} \leq \pi)$
represents the support of $\rho^{(I)}(\theta)$.

%
%

\section{Agreement of the critical temperature}
\label{section:crit-temp}

In this section we show analytically that
the critical temperature of the PWMM
agrees with that of ${\cal N}=4$ SYM
on $R\times S^3$.

Let us first take the $n\rightarrow\infty$ limit
in (\ref{limit}).
Then, the single-particle partition functions
(\ref{zsp}), (\ref{zvp}) and (\ref{zfp}) reduce to
\begin{align}
z_{s}^{(I,J)}(x)
=&x\frac{\partial}{\partial x}\left(
\frac{x^{|I-J|+1}}{1-x^2}\right),
\label{zsp 2} \\
z_{v}^{(I,J)}(x)=&x^2\frac{\partial}{\partial x}
\left(\frac{x^{|I-J|-1+2\delta_{IJ}}}
{1-x^2}\right)
+\frac{\partial}{\partial x}
\left(\frac{x^{|I-J|+3}}
{1-x^2}\right),
\label{zvp 2} \\
z_{f}^{(I,J)}(x)=&x^{\frac{3}{2}}\frac{\partial}{\partial x}
\left(\frac{x^{|I-J|+2\delta_{IJ}}}
{1-x^2}\right)
+x^{\frac{1}{2}}\frac{\partial}{\partial x}
\left(\frac{x^{|I-J|+2}}
{1-x^2}\right) \ .
\label{zfp 2}
\end{align}
Note here that $z_{i}^{(I,J)}(x) \;\;(i=s,v,f)$
and hence $\tilde{V}^{(I,J)}_p$ are $\nu \times \nu$ Toeplitz matrices; i.e.,
\begin{align}
z_{i}^{(I+1,J+1)}(x)=&z_{i}^{(I,J)}(x) \;\;\;\;\mbox{for $i=s,v,f$} \ , \\
\tilde{V}^{(I+1,J+1)}_p=&\tilde{V}^{(I,J)}_p \ .
\end{align}
Therefore, we can represent them by
$z_{i}^{(I,J)}(x)=z_{i}^{(I-J)}(x)$,
$\tilde{V}^{(I,J)}_p=\tilde{V}^{(I-J)}_p$ and
make a Fourier transformation as
\begin{align}
&\hat{z}_{i}(x,\lambda)=
\sum_{K=-\infty}^{\infty}z_{i}^{(K)}(x) \,  e^{iK\lambda}
\;\;\;\; \mbox{for $i=s,v,f$}  \ ,
\label{Fourier transform0}
\\
&\hat{V}_p(\lambda)=\sum_{K=-\infty}^{\infty}
\tilde{V}^{(K)}_p \, e^{iK\lambda}
=\frac{1}{p}
\Bigl\{ 1-6\hat{z}_{s}(x^p,\lambda)
-\hat{z}_{v}(x^p,\lambda)-4(-1)^{p+1}\hat{z}_{f}(x^p,\lambda)
\Bigr\} \ .
\label{Fourier transform}
\end{align}
Let $\tilde{v}^{(I)}_p$ be the eigenvalues of $\tilde{V}^{(I,J)}_p$.
Then, a theorem for a Toeplitz matrix \cite{Gray} implies
\begin{align}
\lim_{\nu\rightarrow\infty}\min_I \tilde{v}^{(I)}_p
=\min_{\lambda} \hat{V}_p(\lambda)
\label{Ttheorem}
\end{align}
for each $p$.
Since $\hat{V}_p(\lambda)$ is positive in the low temperature
phase, (\ref{Ttheorem}) allows us to replace
the condition (\ref{condition for critical temperatute}) by
\begin{align}
\min_{\lambda} \hat{V}_p(\lambda)=0 \quad
\mbox{for certain $p$} \ .
\label{condition for critical temperatute 2}
\end{align}
This can be further replaced by
\begin{align}
\min_{\lambda,\;p} \Bigl\{ p\hat{V}_p(\lambda) \Bigr\} = 0  \ .
\label{condition for critical temperatute 3}
\end{align}
In the appendix \ref{sec:appendixA} we show that the left-hand side
of this equation is given by
\begin{align}
\min_{\lambda,\;p}
\Bigl\{ p\hat{V}_p(\lambda) \Bigr\}
=\hat{V}_1(0)=\tilde{V}_1 \ ,
\label{pcheckV}
\end{align}
where $\tilde{V}_1$ is defined in (\ref{Vdef}).
Therefore, the condition
(\ref{condition for critical temperatute})
is indeed equivalent to
(\ref{condition}),
which implies that the critical temperature should
agree with ${\cal N}=4$ SYM on $R\times S^3$.

Let us comment that eqs.\ (\ref{zsp 2}),(\ref{zvp 2}) and (\ref{zfp 2})
actually coincide with the single-particle partition functions
for $\mathcal{N}=8$ SYM on $R\times S^2$
around a specific monopole background.
{}What we have seen in the previous paragraph
can therefore be
understood as the large-$N$ equivalence between
${\cal N}=4$ SYM on $R\times S^3$
and $\mathcal{N}=8$ SYM on $R\times S^2$ around the corresponding
multi-monopole background \cite{Ishii:2008ib}.
The agreement of the critical temperature for these two theories
has been observed numerically in an earlier work \cite{Grignani:2007xz}.
%
%

\section{Agreement of the free energy}
\label{section:free-energy}

In this section we
demonstrate the agreement of the free energy
(\ref{agreement of free energy}).
%
%
For that purpose
we pay attention to
the single partition functions that appear in the kernel
(\ref{Vij-def})
of the effective action.
We find that $z_{i}^{(I,J)}(x)$ for the PWMM defined in
eqs.\ (\ref{zsp})$\sim$(\ref{zfp})
and $z_{i}(x)$ for the SYM defined
in eq.\ (\ref{single-particle partition function})
are related to each other as
\begin{align}
\label{z-z-rel}
&\sum_{J=1}^{\nu}z_{i}^{(I,J)}(x)=z_i(x)+\Delta z_{i}^{(I)}(x)
\quad\quad
\mbox{for $i=s,v,f$} \ , \\
\mbox{where~~~}
&\Delta z_{s}^{(I)}(x)=
-x\frac{\partial}{\partial x}
\left\{\frac{x}{1-x^2}\Delta^{(I)}(x)\right\} \ , \\
&\Delta z_{v}^{(I)}(x)=
-x^2\frac{\partial}{\partial x}
\left\{\frac{x^{-1}}{1-x^2}\Delta^{(I)}(x)\right\}
-\frac{\partial}{\partial x}
\left\{\frac{x^3}{1-x^2}\Delta^{(I)}(x)\right\} \ ,\\
&\Delta z_{f}^{(I)}(x)=
-x^{\frac{3}{2}}
\frac{\partial}{\partial x}\left\{\frac{1}{1-x^2}\Delta^{(I)}(x)\right\}
-x^{-\frac{1}{2}}
\frac{\partial}{\partial x}\left\{\frac{x^2}{1-x^2}\Delta^{(I)}(x)\right\}  \ ,\\
&\Delta^{(I)}(x)=\frac{1}{1-x}
\Bigl\{
x^{I}+x^{\nu-I+1}+x^{2n-\nu+I}(1-x^{\nu})
\Bigr\} \ .
\end{align}
For instance, (\ref{z-z-rel}) for the $i=s$ case can be shown as
\begin{align}
\sum_{J=1}^{\nu}z_{s}^{(I,J)}(x)
=x\frac{\partial}{\partial x}\left\{\frac{x}{1-x^2}
\sum_{J=1}^{\nu}(x^{|I-J|}-x^{2n-\nu-1+I+J})\right\}
=z_s(x)+\Delta z_{s}^{(I)}(x) \ .
\end{align}
Using (\ref{z-z-rel}), we find that
\begin{align}
&\sum_{J=1}^{\nu}V^{(I,J)}(\theta)
=V(\theta)+\Delta V^{(I)}(\theta) \ ,  \\
&\Delta V^{(I)}(\theta)=
-\sum_{p=1}^{\infty}\frac{1}{p}
\Bigl\{ 6\Delta z^{(I)}_{s}(x^p)+\Delta z^{(I)}_{v}
+4(-1)^{p+1}\Delta z^{(I)}_{f}(x^p) \Bigr\} \cos (p\theta) \ ,
\end{align}
where $V(\theta)$ is the kernel (\ref{Vthetadef})
for the ${\cal N}=4$ SYM,
and the remaining $I$-dependent part
$\Delta V^{(I)}(\theta)$
decreases exponentially as one moves away from
the edges $I=1$ and $I=\nu$.

Thus we may naturally expect
the solution
to (\ref{saddle-point equation for matrix model})
to be
\begin{align}
\rho^{(I)}(\theta)=\hat{\rho}(\theta)+\Delta\rho^{(I)}(\theta) \ ,
\label{solution to saddle-point equation}
\end{align}
where $\hat{\rho}(\theta)$ is the solution
for the ${\cal N}=4$ SYM,
and $\Delta\rho^{(I)}(\theta)$ decreases exponentially
as one moves away from the edges.
We will see such a behavior explicitly by Monte Carlo simulation
in section \ref{section:MCsim}.
Thus we have
\begin{align}
\frac{1}{\nu}\sum_{I=1}^{\nu}|\Delta\rho^{(I)}(\theta)|
= {\cal O}\Bigl(\frac{1}{\nu}\Bigr) \ .
\label{condition for rho}
\end{align}
Substituting (\ref{solution to saddle-point equation})
into (\ref{effective action for matrix model})
and using (\ref{condition for rho}),
we obtain
\begin{align}
\frac{1}{\nu k^2} S_{\rm eff}
= \int
d\theta \, d\theta' \,
\hat{\rho}(\theta) \, V(\theta-\theta') \, \hat{\rho}(\theta')
+ {\cal O}\Bigl(\frac{1}{\nu}\Bigr)  \ ,
\end{align}
and hence the relationship (\ref{agreement of free energy})
in the limit (\ref{limit}).\footnote{In
deriving (\ref{agreement of free energy}),
we did not use the
condition $n-\nu/2 \rightarrow\infty$ in (\ref{limit}),
which amounts to requiring
the smallest fuzzy sphere to be regarded as a
continuous sphere.
It could therefore be that the equivalence holds irrespective of
how one takes the large-$n$ and large-$\nu$ limits as far as
$n-\frac{\nu+1}{2}\geq 0$ is satisfied.}

\FIGURE{
\epsfig{file=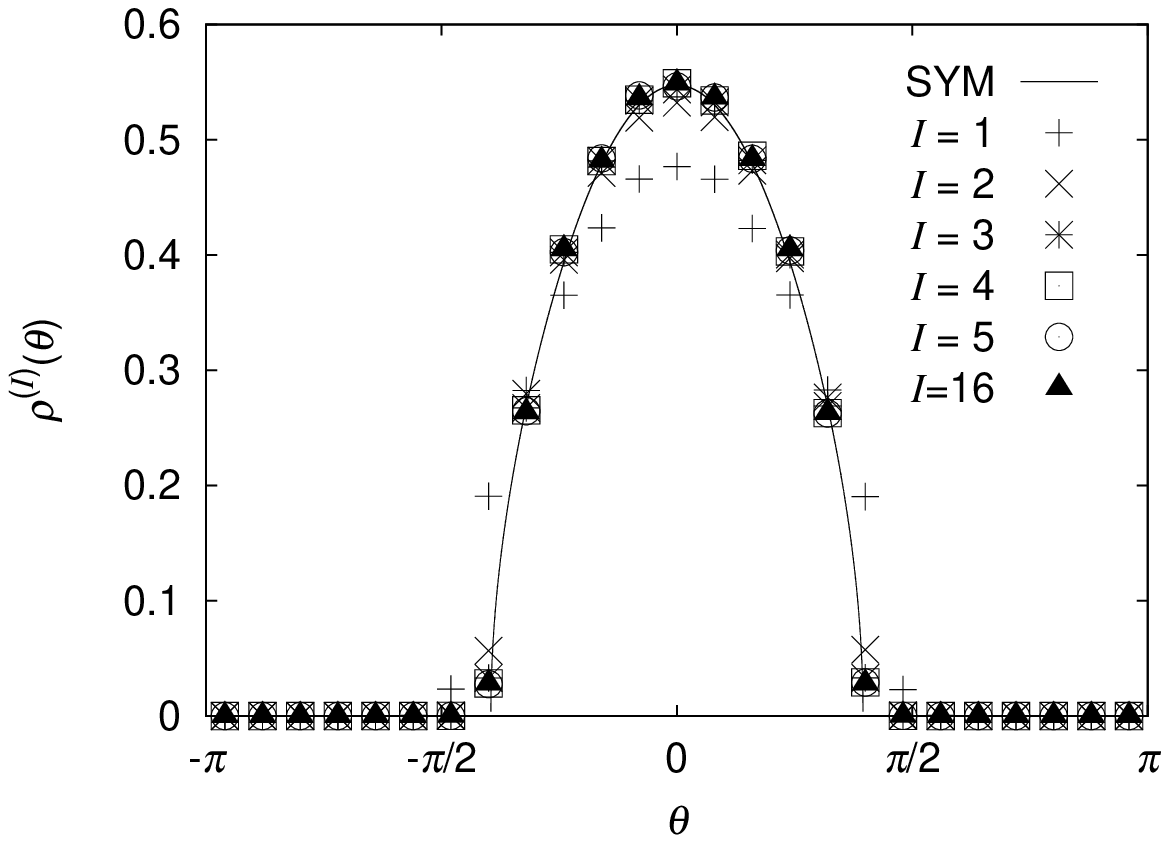,width=10.0cm}
\caption{The distribution $\rho^{(I)}(\theta)$
of the gauge field moduli in
the PWMM around the background (\ref{our background})
is plotted for $I=1,2,3,4,5,16$ with $k=16$ and $n=\nu=31$
at a temperature corresponding to $x=0.104$
near
the critical point $x_c \simeq 0.072$.
(A similar behavior is obtained for decreasing $I=31,30, \cdots, 16$.)
The statistical errors are omitted since they
are smaller than the symbol size.
The solid line represents the result (\ref{eigen_anal})
for the ${\cal N}=4$ SYM on $R \times S^3$ at the same temperature.
}
\label{fig:distribution}
}

\section{Monte Carlo simulation of the effective theory}
\label{section:MCsim}

In this section we confirm the large-$N$ reduction
by performing Monte Carlo simulation of the effective theory
(\ref{effective action for general background})
for the gauge field moduli
with the particular background (\ref{our background}).
For the simulation details including the adopted algorithm,
we refer the reader to ref.\ \cite{Kawahara:2006hs},
where the effective theory
(\ref{effective action for general background})
for simpler backgrounds was studied.
In fig.\ \ref{fig:distribution}
we plot the distribution $\rho^{(I)}(\theta)$ of the gauge field moduli
near the critical temperature for various $I$.
As one goes towards the midpoint $I=(\nu+1)/2$,
the distribution converges rapidly to
the result (\ref{eigen_anal}) for the ${\cal N}=4$ SYM on $R \times S^3$
at the same temperature.

%
%

Thus we have seen that
the distribution of the
gauge field moduli
has the property described below
(\ref{solution to saddle-point equation}).
From the argument in the previous section,
this implies that the free energy of the PWMM
agrees with that of SYM
as in eq.\ (\ref{agreement of free energy}).
In order to confirm the agreement more explicitly,
we calculate the free energy
by using (\ref{def-free-energy}), where
$S_{\rm eff}$ is now replaced by
the expectation value of the effective action
(\ref{effective action for general background})
obtained by the Monte Carlo simulation


We have performed simulations for
$k=40,100,200$ and $n=\nu=15,23,31$
at $x=0.071, 0.074, 0.077$, which are near the
critical point $x_c \simeq 0.072$ .
We first extrapolate our results to $k=\infty$.
As is explained in ref.\ \cite{Kawahara:2006hs},
the leading finite $k$ effect for the normalized free energy
is given by
$\frac{\log k}{k}$, which comes from the first term
of (\ref{Vandermonde}).
In fig.\ \ref{fig:fitKN} (left)
we plot our Monte Carlo results for
the normalized free energy
against $\frac{\log k}{k}$ for $n=\nu=31$.
One can see that our data can be nicely fitted by a
straight line.
This allows us to make an extrapolation to $k=\infty$.
In fig.\ \ref{fig:fitKN} (right) we plot
the extrapolated values against $1/n$,
and find that they lie on a straight line.
This allows us to extrapolate our results to $n=\infty$.
Thus we obtain the normalized free energy
in the limit (\ref{limit}), which is plotted in
fig.\ \ref{fig:free-energy}.
Our results obtained from the PWMM
agree nicely with the known result (\ref{F_SYM})
for the ${\cal N}=4$ SYM on $R \times S^3$.






\FIGURE{
    \epsfig{file=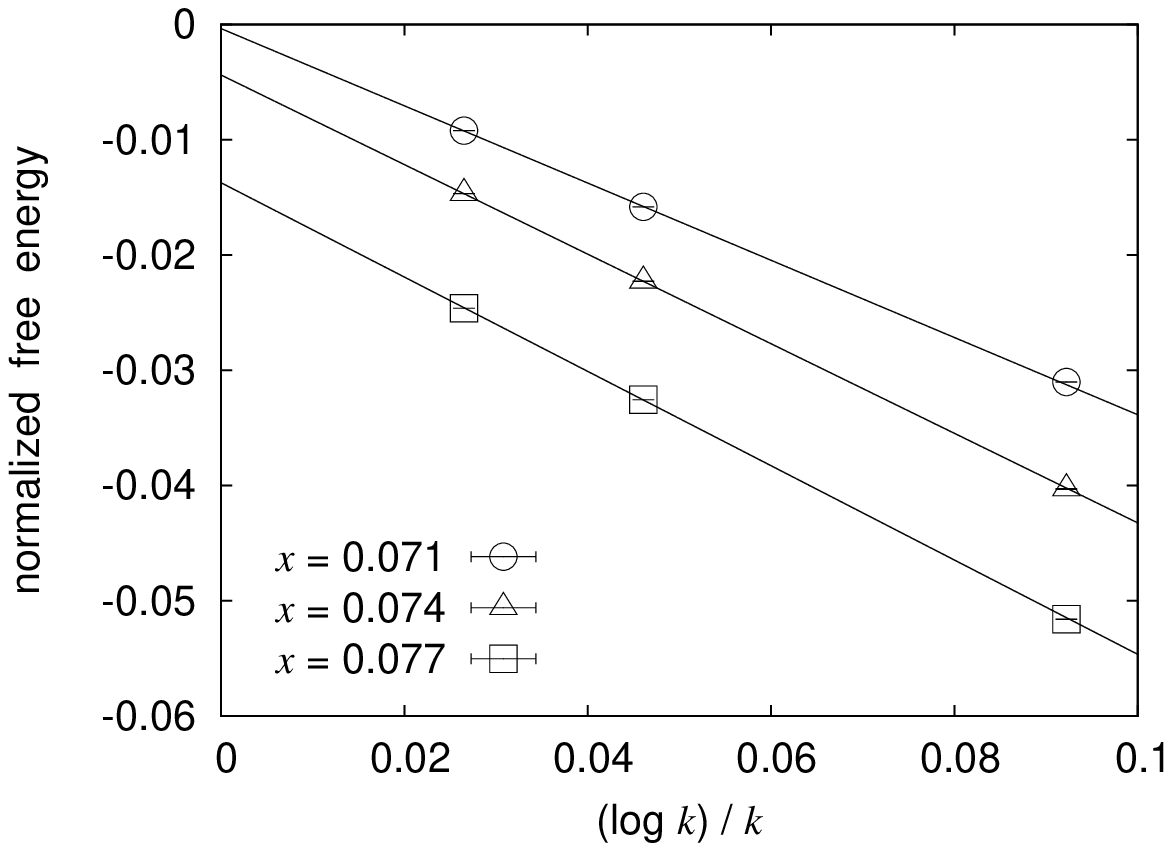,%
width=7.0cm}
    \epsfig{file=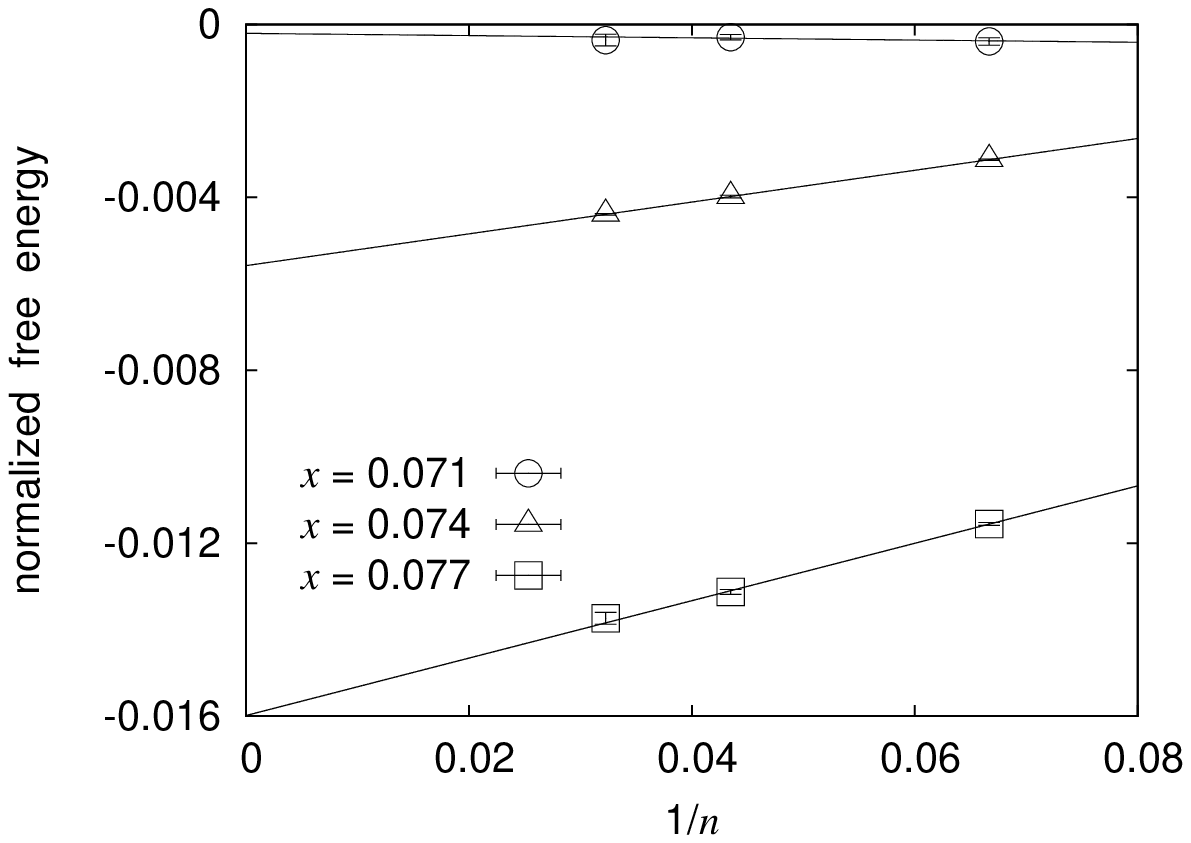,%
width=7.0cm}
\caption{(Left) The normalized free energy
is plotted against $\frac{\log k}{k}$
for $n=\nu=31$.
The error bars represent the statistical error.
The data can be nicely fitted to a straight line, which
enables us to make a large-$k$ extrapolation.
(Right) The results after large-$k$ extrapolation are plotted
against $1/n$.
The error bars represent the fitting error associated with
the large-$k$ extrapolation.
The data can be nicely fitted to a straight line, which
enables us to make a large-$n$ extrapolation.
}
\label{fig:fitKN}
}

\FIGURE{
    \epsfig{file=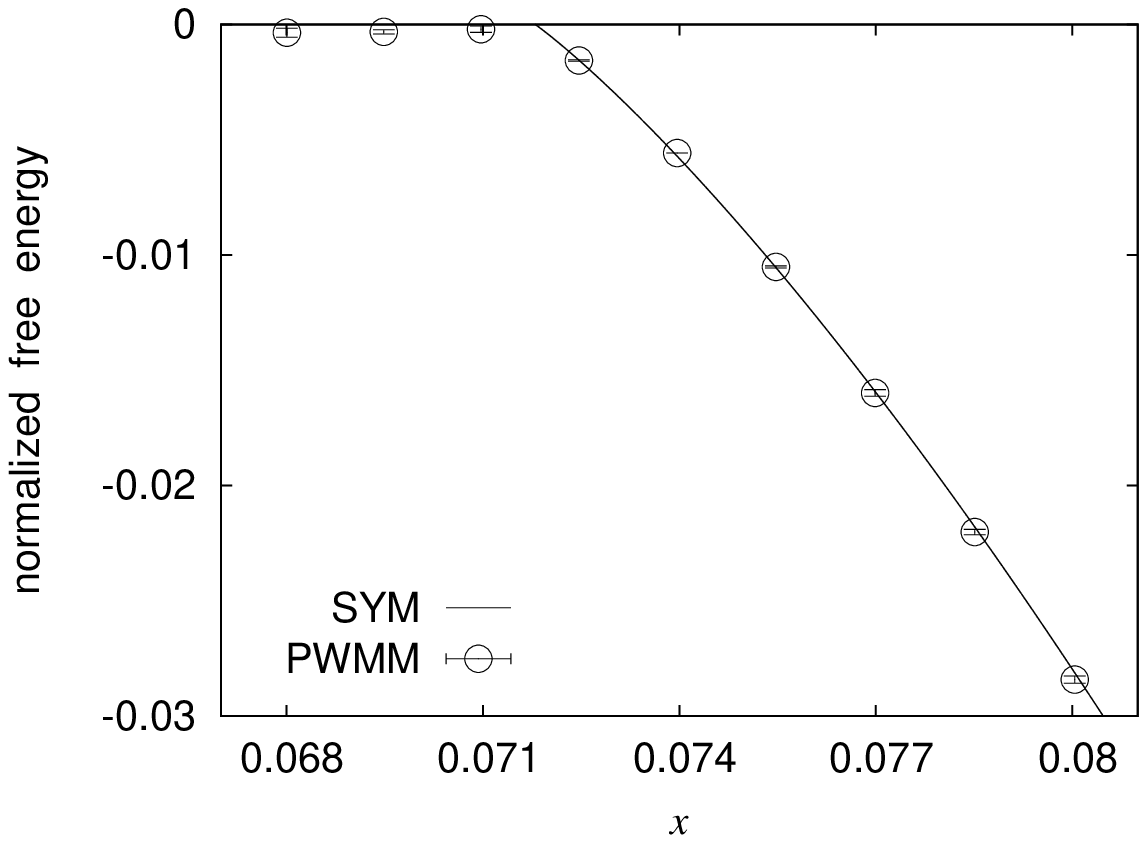,width=10.0cm}
\caption{The normalized free energy
of the PWMM around the background (\ref{our background})
is plotted against the dimensionless parameter $x$ representing
temperature near the critical point $x_c=0.072$.
The data points are obtained by extrapolating results
for $k=40,100,200$ and $n=\nu=15,23,31$.
The error bars represent the fitting error associated with
the large-$n$ extrapolation.
The solid line represents the result (\ref{F_SYM})
for the ${\cal N}=4$ SYM on $R \times S^3$.
}
\label{fig:free-energy}
}

\section{Generalization to $R\times S^3/Z_q$}

Our calculation can be generalized
to $\mathcal{N}=4$ SYM
on $R\times S^3/Z_q$ ($q\neq 1$) \cite{Lin:2005nh}
and $\mathcal{N}=8$ SYM on $R\times S^2$ \cite{Maldacena:2002rb},
from which one obtains the PWMM by shrinking
$S^3/Z_q$ and $S^2$ to a point, respectively.
Unlike the $\mathcal{N}=4$ SYM on $R\times S^3$,
these SYM theories have only ${\rm SU}(2|4)$ supersymmetry.
They have many classical vacua, all of which preserve
this symmetry.
The gravity dual corresponding
to each vacuum is proposed in ref.\ \cite{Lin:2005nh}.
The large-$N$ reduction
is expected to work for each
vacuum by choosing the corresponding background of the PWMM
as one can see from
refs.\ \cite{Ishii:2008ib,Ishiki:2006yr,Ishii:2007ex}.
This regularization preserves the full
${\rm SU}(2|4)$ supersymmetry of the original theory for each vacuum.
In what follows we present the calculation
for $\mathcal{N}=4$ SYM on $R\times S^3/Z_q$,
which reduces to what we have seen above
by setting $q=1$.
A simpler case of $\mathcal{N}=8$ SYM on $R\times S^2$ is
discussed in appendix \ref{sec:appendixB}.
In both cases we restrict ourselves to the trivial vacuum
for simplicity.

Let us consider the theory (\ref{pp-action})
around the background (\ref{background})
in the case
\begin{align}
k_I=k, \;\; n_I = n + q \left(I- \frac{\nu +1}{2} \right)
\quad \quad \mbox{for $I=1,\cdots,\nu $} \ ,
\label{background for S^3/Zq}
\end{align}
and take the large-$N$ limit in such a way
that\footnote{Strictly speaking, one has to tune $\lambda$
as a function of the UV cutoff parameters $n$ and $\nu$
according to the coupling constant renormalization
in the non-conformal case $q \neq 1$.
This issue is irrelevant, though, in the weak coupling limit
we are discussing.\label{foot:renorm}}
\begin{align}
& n \rightarrow \infty, \;\; \nu \rightarrow \infty , \;\;
k \rightarrow \infty , \;\; n-\frac{q\nu}{2} \rightarrow \infty
\nonumber\\
& \mbox{with} \;\;
\lambda \equiv
\frac{g^2 k}{n}
\;\; \mbox{fixed} \ .
\label{continuum limit for RxS3/Zq}
\end{align}
The resulting theory is expected to be equivalent to
the planar limit of $\mathcal{N}=4$ SYM on
$R\times S^3/Z_q$ \cite{Ishii:2008ib,Ishiki:2006yr,Ishii:2007ex}
with the radius of $S^3$
and the 't Hooft coupling constant given, respectively, by
\begin{align}
R_{S^3}=\frac{2}{\mu} \ ,
\label{defRS3-Z}  \quad \quad
\lambda_{R\times S^3/Z_q} = \lambda \, V_{S^3/Z_q} \ ,
\end{align}
where
$V_{S^3/Z_q} = 2\pi^2 (R_{S^3})^3/q$ is the volume of $S^3/Z_q$.
In what follows we confirm this statement
in the weak coupling limit at finite temperature.
The effective action
is given by the same form as
(\ref{effective action for matrix model}),
where $n_I$ in $V^{(I,J)}$ is given by
(\ref{background for S^3/Zq}).

Let us repeat the calculation in section \ref{section:crit-temp}
for general $q$.
We first take the $n \rightarrow \infty$ limit
in (\ref{continuum limit for RxS3/Zq}). Then,
(\ref{zsp}), (\ref{zvp}) and (\ref{zfp}) are reduced to
\begin{align}
z_{s}^{(I,J)}(x)
=&x\frac{\partial}{\partial x}\left(
\frac{x^{q|I-J|+1}}{1-x^2}\right),
\label{zsp zq} \\
z_{v}^{(I,J)}(x)=&x^2\frac{\partial}{\partial x}
\left(\frac{x^{q|I-J|-1+2\delta_{IJ}}}
{1-x^2}\right)
+\frac{\partial}{\partial x}
\left(\frac{x^{q|I-J|+3}}
{1-x^2}\right),
\label{zvp zq} \\
z_{f}^{(I,J)}(x)=&x^{\frac{3}{2}}\frac{\partial}{\partial x}
\left(\frac{x^{q|I-J|+2\delta_{IJ}}}
{1-x^2}\right)
+x^{\frac{1}{2}}\frac{\partial}{\partial x}
\left(\frac{x^{q|I-J|+2}}
{1-x^2}\right) \ .
\label{zfp zq}
\end{align}
Since these are Toeplitz matrices, we can set
$z^{(I,J)}_{i}(x) \equiv z^{(I-J)}_{i}(x)$ ($i=s,v,f$).
The Fourier transforms
(\ref{Fourier transform0}) of these functions
are evaluated as
\begin{align}
\hat{z}_{s}(x,\lambda)&=
\frac{x(1+x^2)(1-x^{2q})}{(1-x^2)^2(1+x^{2q}-2x^qu)}+
\frac{2qx^{q+1}(u-2x^q+ux^{2q})}{(1-x^2)(1+x^{2q}-2x^qu)^2} \ ,
\label{ft of scalar}
\\
\hat{z}_{v}(x,\lambda)&=
\frac{x^2(1+x^2)}{(1-x^2)^2}
+\frac{2x^q(u-x^q)(q-1-(q-3)x^2)+x^2(3-x^2)(1-x^{2q})}
{(1-x^2)^2(1+x^{2q}-2x^qu)} \nonumber\\
& \;\;\;\;
+\frac{2qx^{2q}(2u^2-1-2ux^q+x^{2q})+2qx^{q+2}(u-2x^q+ux^{2q})}
{(1-x^2)(1+x^{2q}-2x^qu)^2} \ ,
\label{ft of vector}
\\
\hat{z}_{f}(x,\lambda)&=
\frac{2x^{\frac{5}{2}}}{(1-x^2)^2}
+\frac{2x^{q+\frac{1}{2}}(u-x^q)(2x^2+q(1-x^2))
+2x^{\frac{3}{2}}(1-x^{2q})}{(1-x^2)^2(1+x^{2q}-2x^qu)}
\nonumber\\
& \;\;\;\;
+\frac{2qx^{2q+\frac{1}{2}}(2u^2-1-2ux^q+x^{2q})
+2qx^{q+\frac{3}{2}}(u-2x^q+ux^{2q})}
{(1-x^2)(1+x^{2q}-2x^qu)^2} \ ,
\label{ft of fermion}
\end{align}
where $u=\cos \lambda$.
One can easily see that
\begin{align}
&\max_{\lambda}\hat{z}_{s}(x,\lambda)=
\hat{z}_{s}(x,0)=
\frac{x(1+x^2)(1-x^{2q})+2qx^{q+1}(1-x^2)}{(1-x^2)^2(1-x^q)^2} \ ,
\nonumber\\
&\max_{\lambda}\hat{z}_{v}(x,\lambda)=
\hat{z}_{v}(x,0)=
\frac{4x^2(1-x^{2q})-2x^q(1-x^2)^2(1-x^q)+2qx^q(1-x^4)}
{(1-x^2)^2(1-x^q)^2} \ ,
\nonumber\\
&\max_{\lambda}\hat{z}_{f}(x,\lambda)=
\hat{z}_{f}(x,0)=
\frac{2x^{\frac{1}{2}}(1+x)(x(1-x^{2q})+qx^q(1-x^2))}
{(1-x^2)^2(1-x^q)^2} \ ,
\label{max of z for S^3/Z_q}
\end{align}
where the right-hand sides
coincide with the single-particle partition functions in SYM on
$R\times S^3/Z_q$ obtained at one loop in ref.\ \cite{Hikida:2006qb}.
By applying the same argument as in appendix \ref{sec:appendixA},
we find\footnote{For general $q$, we had to assume that
the critical point lies in the regime $x \ll 1$
to derive (\ref{min V for S^3/Z_q}) analytically.}
\begin{align}
\min_{\lambda,\;p} \Bigl\{p \hat{V}_p(\lambda) \Bigr\}
=\hat{V}_1(0) \ .
\label{min V for S^3/Z_q}
\end{align}
Therefore the critical temperature
is determined by $\hat{V}_1(0)=0$, which coincides
with the condition for
${\cal N}=4$ SYM on $R\times S^3/Z_q$.
Thus we find that the critical temperature in
${\cal N}=4$ SYM on $R\times S^3/Z_q$ is reproduced
from PWMM.

We can evaluate the free energy with an assumption made
in section \ref{section:free-energy} and see
that it agrees with the continuum theory.
We have also performed Monte Carlo simulations
of the effective theory for the gauge field moduli
as we did for the $q=1$ case and confirmed the assumption
as well as the agreement of the free energy explicitly.

Let us see what happens if we omit fermions and
consider a bosonic theory.
If we introduce $r$ adjoint scalars instead of 6,
the effective action takes
the same form as (\ref{effective action for matrix model})
except that
(\ref{Vtilde}) has to be replaced by
\begin{align}
\tilde{V}^{(I,J)}_p =
\frac{1}{p}
\Bigl\{
\delta_{IJ}-rz^{(I,J)}_{s}(x^p)-z^{(I,J)}_{v}(x^p)  \Bigr\} \ .
\label{Vtilde for S^3/Z_q}
\end{align}
Correspondingly, (\ref{Fourier transform}) has to be replaced by
\begin{align}
\hat{V}_p(\lambda)=\sum_{K=-\infty}^{\infty}\tilde{V}^{(K)}_pe^{iK\lambda}
=\frac{1}{p}
\Bigl\{
1-r\hat{z}_{s}(x^p,\lambda)-\hat{z}_{v}(x^p,\lambda) \Bigr\} \ ,
\label{bosonic V}
\end{align}
where $\hat{z}_{s}(x^p,\lambda)$ and
$\hat{z}_{v}(x^p,\lambda)$ are given by
(\ref{ft of scalar}) and (\ref{ft of vector}), respectively.
From (\ref{max of z for S^3/Z_q}), we find that
(\ref{min V for S^3/Z_q}) holds also for (\ref{bosonic V}).
Therefore, we see that the critical temperature
of the bosonic YM theory on $R\times S^3/Z_q$
is reproduced from the corresponding bosonic matrix model.
We have also checked the agreement of the free energy.

\section{Summary and discussions}
\label{sec:concl-discuss}

In this paper we have provided a nontrivial
test of the large-$N$ reduction for
${\cal N}=4$ SYM on $R\times S^3$, which enables
us to study the AdS/CFT correspondence from first principles.
By expanding the PWMM around a background
corresponding to $R\times S^3$, we reproduced the deconfinement
phase transition in ${\cal N}=4$ SYM on $R\times S^3$
in the planar limit at weak coupling.
The planar ${\cal N}=4$ SYM on $R\times S^3$ can thus be regularized
by the PWMM around that background.
This regularization preserves the ${\rm SU}(2|4)$ supersymmetry,
which is expected to enhance to the full
superconformal ${\rm SU}(2,2|4)$ symmetry in the large-$N$ limit.
Considering that we are dealing with a finite temperature set-up,
which breaks supersymmetry,
our results actually suggest that
the SO(4) symmetry,\footnote{Restoration of the SO(4) symmetry
was also checked by explicit one-loop calculations
at zero temperature \cite{Ishii:2008ib}.
}
which is the bosonic subgroup of the superconformal symmetry,
is restored in the large-$N$ limit.

By choosing a different classical vacuum of the same model,
we also reproduced the deconfinement transition in
SYM on other space-times such as
$R \times S^3/Z_q$ and $R \times S^2$.
These theories only have ${\rm SU}(2|4)$ supersymmetry, which is fully
preserved by our regularization.
The gravity duals are known \cite{Lin:2005nh},
and if one can calculate various quantities on the gravity side,
one can study the gauge-gravity duality from first principles.
It is interesting that the PWMM can be used to regularize
all these theories in the planar limit by just changing the background
configuration.


We were also able to reproduce bosonic theories on
curved spaces from the PWMM without fermions
around the corresponding backgrounds.
However,
we consider that supersymmetry is needed to
protect the classical backgrounds against radiative corrections
at strong coupling.
Therefore, we consider that
one has to choose sufficiently
low temperature or to compactify the Euclidean time direction
supersymmetrically in order for the large-$N$ reduction
to work at strong coupling.

Monte Carlo simulation of the PWMM can be done in
exactly the same way as in the case of D0-brane system,
which simply corresponds to the $\mu=0$ case of the PWMM.
In particular, in order to respect supersymmetry maximally,
we consider it important
not to discretize the Euclidean time direction,
but to use finite numbers of Fourier modes after
an appropriate gauge fixing \cite{Hanada:2007ti,AHNT}.
Indeed in the case of D0-brane system,
the gauge-gravity duality has been confirmed with high precision
including $\alpha '$ corrections \cite{Hanada:2008gy,Hanada:2008ez}.

The novel large-$N$ reduction discussed in this paper
enables us to extend these numerical works
to ${\cal N}=4$ SYM on $R\times S^3$
in a straightforward manner.
One only has to deform the one-dimensional
gauge theory by the mass parameter $\mu$,
and prepare an appropriate initial configuration to obtain
$R\times S^3$ in the large-$N$ limit.
We consider it remarkable that there exists a seemingly feasible
way to simulate ${\cal N}=4$ SYM on $R\times S^3$
at strong coupling,
and hence to investigate the AdS/CFT correspondence
from first principles.
Note, in particular,
that the existing checks of the AdS/CFT correspondence
are restricted to quantities protected by supersymmetry somehow.
We hope that our formulation enables us to calculate
quantities on the gauge theory side without such restrictions
and to compare the results against predictions from the gravity
side.

\acknowledgments

We would like to thank H.\ Kawai, Y.\ Kitazawa and K.\ Matsumoto
for valuable discussions.
The work of G.\ I.\
is supported
by JSPS.
The work of S.-W.\ K.\
is supported by the Center for Quantum Spacetime of Sogang University
(Grant number: R11-2005-021).
The work of J.\ N.\ and A.\ T.\
is supported
by Grant-in-Aid for Scientific
Research (Nos.\ 19340066, 20540286 and 19540294)
from Japan Society for the Promotion of Science.

\appendix
\section{Derivation of eq.\ (\ref{pcheckV})}
\label{sec:appendixA}

In order to
derive
eq.\ (\ref{pcheckV}) in section \ref{section:crit-temp},
let us note first that $\hat{z}_{i}(x,\lambda)$
in (\ref{Fourier transform}) can be written explicitly as
\begin{align}
\hat{z}_{s}(x,\lambda)
&=\frac{x-x^3}{(1+x^2-2xu)^2}
\label{tildezsp} \ , \\
\hat{z}_{v}(x,\lambda)
&=\frac{2x^2(1+2u^2-4xu+x^2)}{(1+x^2-2xu)^2}
\label{tildezvp} \ ,\\
\hat{z}_{f}(x,\lambda)
&=\frac{2x^{\frac{3}{2}}(1-x)(1+u)}{(1+x^2-2xu)^2} \ ,
\label{tildezfp}
\end{align}
where $u=\cos\lambda$. For instance, (\ref{tildezsp}) can be derived as
\begin{align}
\hat{z}_{s}(x,\lambda)
=& x\frac{\partial}{\partial x}
\left(\frac{x}{1-x^2}
\left(1+\sum_{K=1}^{\infty}e^{(\ln x+i\lambda)K}+
\sum_{K=1}^{\infty}e^{(\ln x-i\lambda)K}\right)\right) \\
=& x\frac{\partial}{\partial x}
\left(\frac{x}{1+x^2-2x\cos\lambda}\right)  \ .
\end{align}
Then one can easily find that
\begin{align}
\max_{\lambda}\hat{z}_{i}(x,\lambda)=\hat{z}_{i}(x,0)=z_i(x)
\;\;\;\; \mbox{for $i=s,v,f$} \ ,
\label{maximum of check z}
\end{align}
where
$z_i(x)$ is defined for the $\mathcal{N}=4$ SYM
in eq.\ (\ref{single-particle partition function}).
For odd $p$, this implies that
\begin{align}
\min_{\lambda}
\Bigl\{ p\hat{V}_p(\lambda) \Bigr\}
= p \hat{V}_p(0) = p \tilde{V}_p \ ,
\label{odd-p}
\end{align}
where $\tilde{V}_p$ is given by eq.\ (\ref{Vdef}).
Taking the derivatives of $z_i(x)$, we find
\begin{align}
&\frac{\partial}{\partial x}z_{s}(x)
=\frac{1+4x+x^2}{(1-x)^4} >0 \ , \nonumber\\
&\frac{\partial}{\partial x}z_{v}(x)
=\frac{12x}{(1-x)^4} >0 \ , \nonumber\\
&\frac{\partial}{\partial x}z_{f}(x)
=\frac{6(x^{\frac{1}{2}}+x^{\frac{3}{2}})}{(1-x)^4} >0 \ ,
\label{deriv-z}
\end{align}
which imply $p \tilde{V}_p > \tilde{V}_1$ for odd $p\ne 1$.
%
%
Therefore,
\begin{align}
\min_{\lambda,\;p:{\rm odd}}
\Bigl\{ p\hat{V}_p(\lambda) \Bigr\}
=\hat{V}_1(0)
=\tilde{V}_1 \ .
\label{odd-p2}
\end{align}

The remaining task is to show that
\begin{align}
 p\hat{V}_p(\lambda) > \tilde{V}_1 \quad \quad
\mbox{for even $p$} \ .
\label{final-task}
\end{align}
Let us note first that
\begin{align}
\frac{\partial}{\partial u}(p\hat{V}_p)
=\frac{-24x^p(x^p-x^{3p})+8x^{\frac{3}{2}p}(1-x^p)(1+4x^p+x^{2p})
-8x^{2p}(1-x^p)(1-x^{\frac{1}{2}p})^2u}{(1+x^{2p}-2x^pu)^3} \ ,
\end{align}
from which we find
that $p\hat{V}_p(\lambda)$ is minimized
either at $u=1$ or at $u=-1$ for each $p$.
Next, from (\ref{deriv-z}) we find
that
\begin{align}
\tilde{V}_1 <1-6z_{s}(x^p)-z_{v}(x^p)-4z_{f}(x^p)
=1-\frac{6x^p+12x^{2p}-2x^{3p}+16x^{\frac{3}{2}p}}{(1-x^p)^3}
\equiv F_p \ ,
\end{align}
where we have defined a new function $F_p$ of $x$.
Considering that
\begin{align}
&p\hat{V}_p(0)- F_p=8z_{f}(x^p)>0 \ , \nonumber\\
&p\hat{V}_p(\pi)- F_p
=\frac{16x^p(3x^p+2x^{2p}+3x^{3p}+x^{\frac{1}{2}p}(1+x^p)^3)}
{(1-x^{2p})^3} >0 \ ,
\end{align}
we obtain the inequality (\ref{final-task}).
From (\ref{odd-p2}) and (\ref{final-task}), we obtain (\ref{pcheckV}).

\section{Large-$N$ reduction for $\mathcal{N}=8$ SYM on $R \times S^2$}
\label{sec:appendixB}

In this appendix we consider
$\mathcal{N}=8$ SYM on $R \times S^2$.
The ``large-$N$ reduction'' in this case is nothing but
the well-known construction of planar field theories
using fuzzy spheres.
We discuss it here nevertheless to see how our calculations
reduce in this simpler case.

Let us consider the theory (\ref{pp-action})
around the background (\ref{background}) in the case
\begin{align}
\nu =1, \;\; k_1=k, \;\; n_1 = n \ ,
\label{background for S^2}
\end{align}
and take the large-$N$ limit in such a way that
(See footnote \ref{foot:renorm}.)
\begin{align}
 n \rightarrow \infty, \;\; k \rightarrow \infty \;\;\;
\mbox{with} \;\;
\lambda \equiv  \frac{g^2 k}{n}
\; \; \mbox{fixed} \ .
\label{commutative limit of fuzzy sphere}
\end{align}
The resulting theory is equivalent to the planar limit of
$\mathcal{N}=8$ SYM on $R\times S^2$,
with the radius of $S^2$
and the 't Hooft coupling constant given,
respectively, by\footnote{The relationship between the radius of $S^2$
and the parameter $\mu$ of the PWMM agrees with the one obtained in
dimensionally reducing
${\cal N}=4$ SYM on $R\times S^2$ to arrive at
the PWMM (\ref{pp-action}).
The fact that $R_{S^2}$ is half the radius of $S^3$
in (\ref{defRS3}) can be understood by regarding
$S^3$ as an $S^1$ bundle over $S^2$ and by dimensionally
reducing the $S^1$ fiber direction to obtain $S^2$.
}
\begin{align}
R_{S^2}=\frac{1}{\mu} \ ,
\label{defRS2}  \quad \quad
 \lambda_{R \times S^2} = \lambda \, V_{S^2} \ ,
\end{align}
where $V_{S^2} = 4\pi (R_{S^2})^2$ is the volume of $S^2$.
After taking the limit (\ref{commutative limit of fuzzy sphere}),
we find that (\ref{zsp}), (\ref{zvp}) and (\ref{zfp}) are reduced to
\begin{align}
\label{zsS2}
& z_s^{(1,1)}(x)=\frac{x(1+x^2)}{(1-x^2)^2} \ , \\
\label{zvS2}
& z_v^{(1,1)}(x)=\frac{4x^2}{(1-x^2)^2}  \ ,  \\
\label{zfS2}
& z_f^{(1,1)}(x)=\frac{2x^{3/2}(1+x)}{(1-x^2)^2} \ .
\end{align}
where the dimensionless parameter $x$ is
defined in eq.\ (\ref{defx})
with $R_{S^3} = 2/\mu$.
Rewriting (\ref{zsS2})$\sim$(\ref{zfS2}) in terms of
\begin{align}
\tilde{x} = \exp \left(- \frac{1}{R_{S^2}T}\right)
= \exp \left(- \frac{\mu}{T}\right) = x^2 \ ,
\end{align}
they completely agree with the single-particle partition functions in
$\mathcal{N}=8$ SYM on
$R\times S^2$ obtained at one loop in ref.\ \cite{Grignani:2007xz}.
Therefore, the free energy agrees with that of
$\mathcal{N}=8$ SYM on $R\times S^2$.

We can redo the calculation
omitting fermions
in the PWMM.
The free energy of the resulting bosonic matrix model
around the background (\ref{background for S^2})
agrees
in the limit (\ref{commutative limit of fuzzy sphere})
with the corresponding bosonic theory on $R\times S^2$.

\end{document}